\documentclass[conference]{IEEEtran}
\IEEEoverridecommandlockouts
\usepackage{amsmath,amsfonts}
\usepackage{algorithmic}
\usepackage{algorithm}
\usepackage{array}
\usepackage[font=normalsize,labelfont=sf,textfont=sf]{subcaption}
\usepackage{textcomp}
\usepackage{stfloats}
\usepackage{url}
\usepackage{verbatim}
\usepackage{graphicx}
\usepackage{xcolor}
\usepackage{bbm}
\usepackage{amsthm}
\newfont{\bbb}{msbm10 scaled 700}

\DeclareSymbolFont{bbsymbol}{U}{bbold}{m}{n}

\newfont{\bb}{msbm10 scaled 1100}

\DeclareMathSymbol{\CC}{\mathbin}{bbsymbol}{'103}
\DeclareMathSymbol{\PP}{\mathbin}{bbsymbol}{'120}
\DeclareMathSymbol{\RR}{\mathbin}{bbsymbol}{'122}
\DeclareMathSymbol{\QQ}{\mathbin}{bbsymbol}{'121}
\DeclareMathSymbol{\ZZ}{\mathbin}{bbsymbol}{'132}
\DeclareMathSymbol{\FF}{\mathbin}{bbsymbol}{'106}
\DeclareMathSymbol{\GG}{\mathbin}{bbsymbol}{'107}
\DeclareMathSymbol{\EE}{\mathbin}{bbsymbol}{'105}
\DeclareMathSymbol{\NN}{\mathbin}{bbsymbol}{'116}
\DeclareMathSymbol{\KK}{\mathbin}{bbsymbol}{'113}
\DeclareMathSymbol{\HH}{\mathbin}{bbsymbol}{'110}
\DeclareMathSymbol{\SSS}{\mathbin}{bbsymbol}{'123}
\DeclareMathSymbol{\UU}{\mathbin}{bbsymbol}{'125}
\DeclareMathSymbol{\VV}{\mathbin}{bbsymbol}{'126}
\DeclareMathSymbol{\XX}{\mathbin}{bbsymbol}{'130}
\DeclareMathSymbol{\BB}{\mathbin}{bbsymbol}{'102}

\DeclareMathSymbol{\yy}{\mathbin}{bbsymbol}{'171}
\DeclareMathSymbol{\xx}{\mathbin}{bbsymbol}{'170}
\DeclareMathSymbol{\zz}{\mathbin}{bbsymbol}{'172}
\DeclareMathSymbol{\sss}{\mathbin}{bbsymbol}{'163}
\DeclareMathSymbol{\rr}{\mathbin}{bbsymbol}{'162}
\DeclareMathSymbol{\pp}{\mathbin}{bbsymbol}{'160}
\DeclareMathSymbol{\qq}{\mathbin}{bbsymbol}{'161}
\DeclareMathSymbol{\ww}{\mathbin}{bbsymbol}{'167}
\DeclareMathSymbol{\hh}{\mathbin}{bbsymbol}{'150}
\DeclareMathSymbol{\uu}{\mathbin}{bbsymbol}{'165}
\DeclareMathSymbol{\vvv}{\mathbin}{bbsymbol}{'166}

\DeclareMathSymbol{\indicator}{\mathbin}{bbsymbol}{'061}

\usepackage[mathscr]{euscript}

\newcommand{\hv}{{\bf h}}

\newcommand{\rv}{{\bf r}}

\newcommand{\uv}{{\bf u}}
\newcommand{\wv}{{\bf w}}
\newcommand{\vv}{{\bf v}}

\newcommand{\yv}{{\bf y}}
\newcommand{\zv}{{\bf z}}
\newcommand{\zerov}{{\bf 0}}

\newcommand{\Fm}{{\bf F}}

\newcommand{\Id}{{\bf I}}

\newcommand{\Bc}{{\cal B}}
\newcommand{\Cc}{{\cal C}}

\newcommand{\Ec}{{\cal E}}

\newcommand{\Gc}{{\cal G}}

\newcommand{\Kc}{{\cal K}}
\newcommand{\Lc}{{\cal L}}

\newcommand{\Nc}{{\cal N}}

\newcommand{\Qc}{{\cal Q}}

\newcommand{\Sc}{{\cal S}}

\newcommand{\Uc}{{\cal U}}

\newcommand{\Xc}{{\cal X}}
\newcommand{\Yc}{{\cal Y}}

\newcommand{\nuv}{\hbox{\boldmath$\nu$}}

\newcommand{\phiv}{\hbox{\boldmath$\phi$}}

\newcommand{\Sigmam}{\hbox{\boldmath$\Sigma$}}

\newcommand{\trace}{{\hbox{tr}}}

\newcommand{\eqdef}{\stackrel{\Delta}{=}}

\newcommand{\herm}{{\sf H}}

\newcommand{\SINR}{{\sf SINR}}
\newcommand{\SNR}{{\sf SNR}}

\newcommand{\taudmrs}{\tau_p}

\newcommand{\Ktot}{K_{\rm tot}}

\usepackage{stmaryrd} % for \mkv 

\newcommand{\defeq}{ \stackrel{\triangle}{=} }
\newcommand{\fhratekUL}{B_{\ell, k}}
\newcommand{\fhratekDL}{R^{\rm dl}_k}

\DeclareOldFontCommand{\rm}{\normalfont\rmfamily}{\mathrm}
\DeclareOldFontCommand{\sf}{\normalfont\sffamily}{\mathsf}
\DeclareOldFontCommand{\tt}{\normalfont\ttfamily}{\mathtt}
\DeclareOldFontCommand{\bf}{\normalfont\bfseries}{\mathbf}
\DeclareOldFontCommand{\it}{\normalfont\itshape}{\mathit}
\DeclareOldFontCommand{\sl}{\normalfont\slshape}{\@nomath\sl}
\DeclareOldFontCommand{\sc}{\normalfont\scshape}{\@nomath\sc}

\usepackage[font={small}]{caption}

\allowdisplaybreaks

\usepackage{cite}

\usepackage[left=.635in,right=.635in,top=.71in,bottom=1.02in]{geometry}

\begin{document}
\bstctlcite{IEEEexample:BSTcontrol}

\title{\huge Optimizing Fronthaul Quantization for Flexible User Load \\
in Cell-Free Massive MIMO\\
}
\author{
	\IEEEauthorblockN{Fabian G\"{o}ttsch\IEEEauthorrefmark{1}\IEEEauthorrefmark{2}, 
     Max Franke\IEEEauthorrefmark{1},
     Arash Pourdamghani\IEEEauthorrefmark{1},
            Giuseppe Caire\IEEEauthorrefmark{1}\IEEEauthorrefmark{2} and Stefan Schmid\IEEEauthorrefmark{1}%\IEEEauthorrefmark{3}
	}
\IEEEauthorblockA{\IEEEauthorrefmark{1}Technical University Berlin, Berlin, Germany}
\IEEEauthorblockA{\IEEEauthorrefmark{2}Massive Beams, Berlin, Germany}
% \IEEEauthorblockA{\IEEEauthorrefmark{3}Fraunhofer Institute for Telecommunications Heinrich-Hertz-Institut, Berlin, Germany}
\IEEEauthorblockA{E-mail: \{fabian.goettsch, mfranke, pourdamghani, caire, 	stefan.schmid\}@tu-berlin.de}
}
  
\maketitle
\begin{abstract}
We investigate the physical layer (PHY) spectral efficiency and fronthaul network load of a scalable user-centric cell-free massive MIMO system. Each user-centric cluster processor responsible for cluster-level signal processing is located at one of multiple decentralized units (DUs). Thus, the radio units in the cluster must exchange data with the corresponding DU over the fronthaul. Because the fronthaul links have limited capacity, this data must be quantized before it is sent over the fronthaul. We consider a routed fronthaul network, where the cluster processor placement and fronthaul traffic routing are jointly optimized with a mixed-integer linear program. For different numbers of users in the network, we investigate the effect of fronthaul quantization rates, a system parameter computed based on rate-distortion theory. Our results show that with optimized quantization rates, the fronthaul load is quite stable for a wide range of user loads without significant PHY performance loss. This demonstrates that the cell-free massive MIMO PHY and fronthaul network are resilient to varying user densities.
\end{abstract}


\section{Introduction}
Cell-free massive MIMO is one of the promising technologies to meet the demands of future 6G wireless communications. 
As each user equipment (UE) is served by the joint processing of
spatially distributed infrastructure antennas, i.e., a user-centric cluster of radio units (RUs), the effects of pathloss and blocking are mitigated, while the macrodiversity is increased and cell boundaries are removed. If the system is well designed and each UE can be served by every RU with a significant channel gain, this design results in a system with highly reduced interference compared to cellular networks.

While the physical layer (PHY) has been investigated in countless works, the fronthaul has only recently been considered in few works (see \cite{ngo2024ultradense} and references therein). A scalable fronthaul network is considered in \cite{li2023joint, Joshi2024fronthaul}, where each user-centric cluster processor is located at one of $N$ decentralized units (DUs).\footnote{We use the scalability definition of \cite{bjo2020}, where a system is considered scalable if the communication and computation complexity at each network component remains finite if the network area grows infinitely with constant density of UEs, RUs and DUs.
% If all user-centric clusters are processed by a centralized unit, the system is obviously not scalable.
}  
In \cite{li2023joint, Joshi2024fronthaul}, the maximum fronthaul link load is minimized by routing the fronthaul traffic and placing the cluster processors at one of the $N$ DUs. 
% The fronthaul traffic routing is done via routers \cite{li2023joint} or inter-DU links \cite{Joshi2024fronthaul}. 
The traffic routing is done via dedicated routers placed in the fronthaul network \cite{li2023joint} or routers collocated at DUs, utilizing inter-DU links \cite{Joshi2024fronthaul}.
In these works, a constant user load is considered.

In a practical network however, the number of UEs in the network can vary drastically, e.g., on a university campus or at a sports venue.
By scheduling $K$ active UEs out of the total number $\Ktot$ of UEs in the network on time-frequency resources, the long-term average throughput rate and idle time of each UE can be optimized.
% , where on each resource approximately the same number of UEs is simultaneously active. A relative good trade-off between the physical layer 
% user spectral efficiency (SE) and 
% sum SE of the $K$ active UEs 
% is to serve approximately $LM/2$ UEs simultaneously \cite{gottsch2023fairness}, where $L$ is the number of RUs, each equipped with $M$ antennas.
% Nevertheless, d
Depending on the user demands, it may be more beneficial to serve many users simultaneously at lower data rates than to serve fewer users at high data rates but with longer idle times.
In this case, even with optimized fronthaul routing and cluster placement, increased fronthaul link load is expected due to the larger number of simultaneously active UEs.
However, another possible optimization variable to optimize the fronthaul load is the distortion level $D$ with regard to the quantization accuracy. 

Fronthaul network planners are generally interested in two metrics: the expected average fronthaul load 
% {\BLUE [bandwidth requirement = fronthaul link load? Btw, we always say fronthaul load and not fronthaul rate because it may be irritating to use rate for fronthaul and physical layer. Bandwidth is also an expression I would rather avoid. For network planning this is the data rate, right? For us, it is the amount of spectrum (i.e., how many Hz) is used for communication.]} 
and the worst-case fronthaul load at peak data traffic events (i.e., very high user load). The average load is important because designing and deploying networks only around peak traffic is costly and will leave vast amounts of unused capacity most of the time \cite{chih2015rethink}. 
However, it is also crucial to understand 
what peak traffic events look like due to an increase of demanded user data rates or simultaneously active users in the network.
% Although the first case can be handled by rate limiting and network slicing, the second case is less easy to deal with. 
% If the number of users becomes too large
Then, even PHY data rates for the most basic functions such as text messages or calls are hard to achieve, e.g., during sports games or emergencies \cite{dilmaghani2006designing}. 
% Understanding peak traffic events allows operators to 
% design the fronthaul and RAN
% in such a way that they do not collapse in these scenarios.
%%%%
Understanding peak traffic events allows operators to design the fronthaul and RAN to be resilient so that they do not collapse under such conditions.
% \footnote{Of course, even if the RAN is lightly loaded but the fronthaul network is the bottleneck, PHY data rates will degrade.} 
% It also allows provisions to be made for a fast scaling up, such as assigning additional computation resources to DUs. As traffic prediction has become very sophisticated \cite{joshi2015review}, it is now common to anticipate traffic patterns and prepare accordingly. 

For these reasons, it is important to understand how a relatively small number of users with high data rates will affect the fronthaul network compared to many users with lower data rates. 
% This will enable network planners to make RANs more resilient during peak data traffic events. 
Especially for user-centric cell-free massive MIMO with a routed fronthaul and distributed DUs and RUs, this is a complex and not well-studied problem. 

\subsection{Contributions}
%This is particularly interesting for the design of fronthaul networks, which are generally planned for the worst case, i.e., a very large user load, although this may occur very rarely in reality.
%This is done because ...
%{\BLUE [@Max: could you add some sentences here why the fronthaul is planned for the worst case? And what the benefits are if the same fronthaul can be used for a wide range of different user loads?]}
% Fronthaul network planning is even more complicated in user-centric cell-free massive MIMO, when UEs and RUs are not associated to one of the DUs based on the fronthaul architecture. 
This paper contributes a first study of how the number of simultaneously active UEs $K$ affects the fronthaul load. This is relevant as it will enable network planners to make RANs more
resilient during peak data traffic events. 
% We investigate a cell-free massive MIMO system with a flexible number $K$ of simultaneously active UEs. 
We identify the user load, for which the RAN operates at full PHY performance, and optimize the fronthaul quantization distortion level $D$. 
% We study the impact of $K$ and the quantization distortion level $D$ on the fronthaul link load and on the physical layer performance.
Our results show that the fronthaul load for high user loads is virtually the same as in a lightly loaded system. Further, the PHY spectral efficiency (SE) is degraded only very slightly for optimized $D$ with regard to the fronthaul compared to smaller distortion levels.

% \emph{Notations:} The blackboard bold letters ${\mathbb{E}}$ and ${\mathbb{C}}$ denote the expectation operator and complex number field, respectively; the notations $(\cdot)^{\rm{T}}$ and $(\cdot)^{\rm{H}}$ denote the transpose and the Hermitian transpose for a matrix, respectively; 
% $\textrm{vec}(\cdot)$ denotes the vectorization operation; 
% ${{{\bf{F}}_N}}$ and ${{{\bf{I}}_M}}$ denote the discrete Fourier transform (DFT) matrix of size $N\times N$ and the identity matrix of size $M\times M$; $x[l]$ denotes the $l$-th entry of the vector $\bf x$ and $a_{i,j}$ denotes the $(i,j)$-th element of the matrix $\bf A$.

%%%%%%
%%%%%%
%%%%%%
%%%%%%
%%%%%%
%%%%%%
\section{System Model}
\begin{figure}[t!]
    \centerline{\includegraphics[trim={0 0 0 0}, clip,width=.99\linewidth]{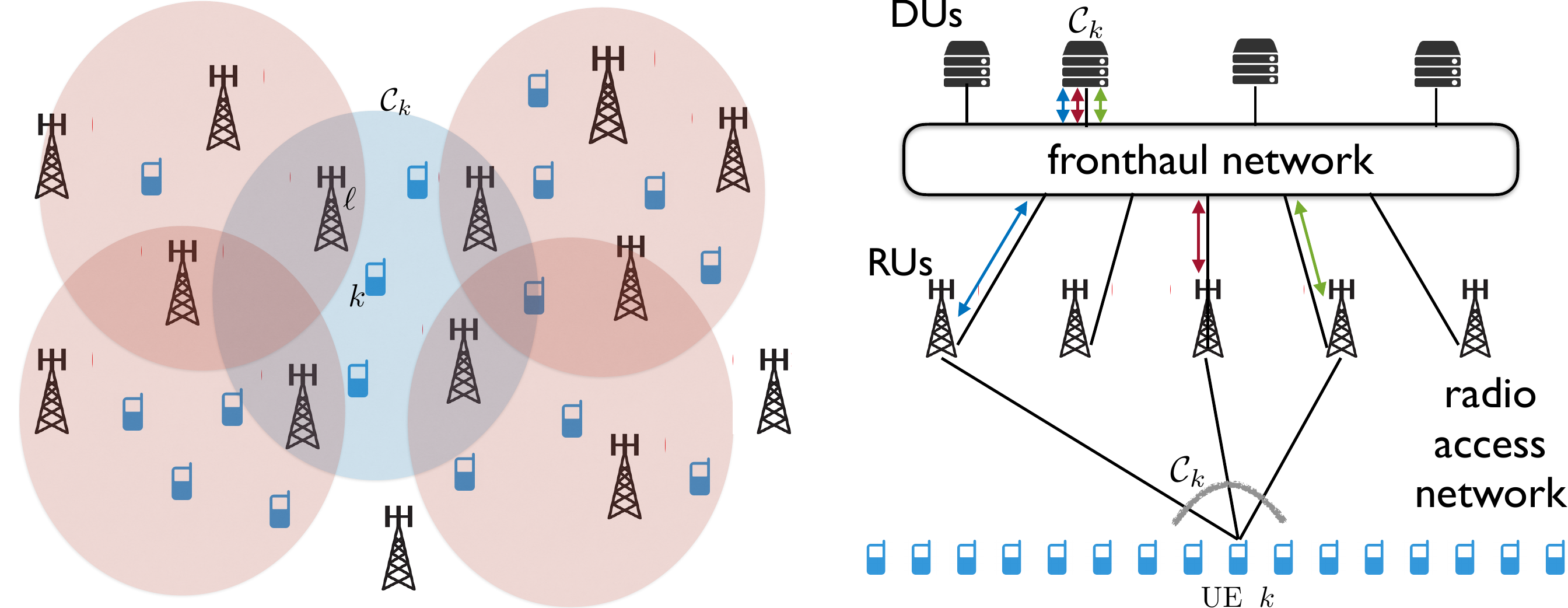}}
    \vspace{-.0cm}
    \caption{
    % An example of user-centric clusters, UE-RU association and the data exchange over the fronthaul. All RUs in $\Cc_k$ are connected through the fronthaul to the DU hosting the cluster processor of $\Cc_k$.
    An example of user-centric clusters, UE-RU association and the data exchange over the fronthaul for user $k$ and its cluster $\Cc_k$.
    }
    \vspace{-17pt}		
    \label{clusters}
\end{figure}
We consider a cell-free massive MIMO RAN and a fronthaul network with a set of $Q$ routers $\mathcal Q=\{1,\dots, Q\}$ and $N$ DUs $\mathcal N=\{1,\dots, N\}$, respectively.
The RAN consists of $K$ single-antenna UEs $\mathcal K=\{1,\dots, K\}$ and $L$ RUs $\mathcal L=\{1,\dots, L\}$, where each RU is equipped with $M$ antennas. The UE-RU associations in the RAN are described by a bipartite graph $\Gc_{\rm ran}(\Kc, \Lc, \Ec_{\rm ran})$ with two classes of nodes (UEs and RUs). The user-centric RU cluster serving user $k \in \Kc$ is denoted by $\Cc_k  \subseteq \Lc$. 
The set of UEs served by RU $\ell \in \Lc$ is denoted by $\Uc_\ell \subseteq \Kc$. The set of edges of $\Ec_{\rm ran}$ is such that $(\ell, k) \in \Ec_{\rm ran}$ iff $\ell \in \Cc_k$ (or, equivalently, iff $k \in \Uc_\ell$).
The fronthaul network connects each RU with a subset of the routers, which are partially connected to the other routers and DUs. 
It is described as a graph $\Gc_{\rm front}(\Lc, \Qc, \Nc, \Ec _{\rm front})$, where the edges $\Ec _{\rm front}$ represent the connections between RUs, routers and DUs. 
The overall network topology is described by the union of $\Gc_{\rm ran}(\Kc, \Lc, \Ec_{\rm ran})$ and $\Gc_{\rm front}(\Lc, \Qc, \Nc, \Ec _{\rm front})$ obtained by merging the common nodes $\Lc$ (see Fig. \ref{clusters}).

The matrix $\HH \in \CC^{LM \times K}$ describes the channels between the $K$ UEs and $LM$ RU antennas. It consists of the vectors $\hv_{\ell,k} \in \CC^{M \times 1}$ representing the channel between UE $k$ and RU $\ell$. Each channel is a correlated complex circularly
symmetric Gaussian vector with mean zero and covariance
matrix
$\Sigmam_{\ell,k} = \EE[ \hv_{\ell,k}\hv_{\ell,k}^\herm]$, 
where we use the well-known 
notation $\hv_{\ell,k} \sim \mathcal{CN} ({\bf 0},\Sigmam_{\ell,k})$.
The large-scale fading coefficient (LSFC) corresponding to $\hv_{\ell, k}$ is defined 
as 
$\beta_{\ell,k} \defeq \frac{1}{M} \trace(\Sigmam_{\ell,k}).$
% \begin{align*}
% \beta_{\ell,k} \defeq \frac{1}{M} \trace(\Sigmam_{\ell,k}).
% \end{align*} 
It describes the average signal attenuation between 
RU $\ell$ and user $k$ due to distance and other macroscopic effects. 
The channel coefficients remain constant over
blocks of $T$ signal dimensions (time-frequency channel uses),
referred to as ``resource blocks'' (RBs). This is a standard model in the (cell-free) massive MIMO literature (see \cite{marzetta2010noncooperative, bjo2020}). For simplicity, we
focus on a generic resource block and we omit the resource block index in this paper.

\subsection{Cluster and pilot allocation}
The user-centric clusters and uplink (UL) pilots for channel estimation are assigned following the subspace information aided overloaded pilot assignment scheme from \cite{osawa2023overloaded}. 
All UEs transmit with the same average energy per symbol $P^{\rm ue}$ and we define the system parameter 
$\SNR \defeq \frac{P^{\rm ue} }{N_0}$,
where $N_0$ denotes the complex baseband noise power spectral density.
Each UE is associated to a cluster $\Cc_k$ of RUs of cardinality at most 
$\Cc_{\rm max}$ (a system parameter) with the largest LSFC, provided that 
\begin{gather}
	\beta_{\ell,k} \geq \frac{\eta}{M \SNR} , \label{eq:snr_threshold}
\end{gather}
where  $\eta > 0$ is a suitable threshold that defines how much larger the useful signal in the presence of 
maximum possible beamforming gain (equal to $M$) should be compared to the noise floor. 
We assume that $\taudmrs$ out of $T$ signal dimensions per RB are dedicated to
pilots and use a codebook of $\taudmrs$ orthogonal pilot 
sequences $\{ \phiv_t : t \in [\taudmrs]\}$. 
Let $\Fm_{\ell,k}$ denote the matrix of the orthonormal eigenvectors of $\Sigmam_{\ell,k}$ spanning 
the {\em dominant channel subspace}, i.e., the subspace containing a sufficiently large fraction of the total channel energy 
$\trace(\Sigmam_{\ell,k}) = M \beta_{\ell,k}$ (e.g., see \cite{goettsch2022,adhikary2013joint}). 
% Then, we say that two users $k,k'$ are approximately mutually orthogonal in the spatial domain with respect to RU $\ell$ if 
% \begin{align*}
% \trace (\Fm_{\ell,k}^\herm \Fm_{\ell,k'} \Fm_{\ell,k'}^\herm \Fm_{\ell,k} ) \leq \epsilon,
% \end{align*} for some fixed threshold $\epsilon$. 
A RU can assign a pilot to multiple UEs under the condition that
the user channel subspaces are approximately mutually orthogonal \cite{li2023joint}.

We let $t_i$ denote the pilot index of UE $i$. Then, each RU $\ell$ obtains an estimate of the channel vectors 
$\{\hv_{\ell,k} : k \in \Uc_\ell\}$ using the subspace projection method of  \cite{goettsch2022}, given by
\begin{align}
	\widehat{\hv}_{\ell,k} = \hv_{\ell,k}  + \Fm_{\ell,k}\Fm_{\ell,k}^\herm \left ( \sum_{ i \neq k : t_i = t_k} \hv_{\ell,i} \right ) + \Fm_{\ell,k} \Fm_{\ell,k}^\herm \widetilde{\zv}^{(t)}_{\ell},   \label{chest1} \nonumber
\end{align}
where $\widetilde{\zv}^{(t)}_{\ell}$ is projected additive white Gaussian noise (AWGN) with i.i.d. components $\sim \Cc\Nc(0, \frac{1}{\taudmrs\SNR})$ and the sum in the second term contains the channels of co-pilot users with respect to UE $k$. 
Since orthogonal pilots are used, the contributions of non-co-pilot users are removed from the received pilot signal by multiplying with UE $k$'s pilot sequence.
% We use the pilot energy normalization
% \begin{align*}
% \| \phiv_{t} \|^2 = \taudmrs \SNR 
% \end{align*} 
% for all $t \in [\taudmrs]$, and let $t_i$ denote the index of the pilot sequence $\phiv_{t_i}$ allocated to user $i$. 
% The pilot field received at RU $\ell$ is given by the $M \times \taudmrs$ matrix 
% \begin{equation} 
% 	\Ym_{\ell}^{\rm pilot} = \sum_{i=1}^K \hv_{\ell,i} \phiv_{t_i}^\herm + \Zm_{\ell}^{\rm pilot}, \label{Y_pilot}
% \end{equation}
% where $\Zm_{\ell}^{\rm pilot}$ is additive white Gaussian noise (AWGN) 
% with elements i.i.d. $\sim \Cc\Nc(0, 1)$.
% Then, RU $\ell$ projects the received pilot field onto a pilot sequence $t \in [\taudmrs]$, obtaining
% \begin{eqnarray} 
% 	\yv^{(t)}_{\ell} & = & \frac{1}{\taudmrs \SNR} \Ym^{\rm pilot}_{\ell} \phiv_{t}  
% 	= \sum_{ i  : t_i = t} \hv_{\ell,i}  + \widetilde{\zv}^{(t)}_{\ell}.  \label{chest}
% \end{eqnarray} 
% Beyond the projected AWGN $\widetilde{\zv}^{(t)}_{\ell}$  with i.i.d. components  $\sim \Cc\Nc(0, \frac{1}{\taudmrs\SNR})$, 
% $\yv^{(t)}_{\ell}$ contains the superposition of the channels of all users $i$ using the pilot $t$. 
% With respect to a given user $k$ with $t_k = t$,  
% the CSI obtained at RU $\ell$ with the subspace projection
% pilot decontamination scheme from \cite{goettsch2022} is given by
% \begin{align}
% 	\widehat{\hv}_{\ell,k} = \hv_{\ell,k}  + \Fm_{\ell,k}\Fm_{\ell,k}^\herm \left ( \sum_{ i \neq k : t_i = t} \hv_{\ell,i} \right ) + \Fm_{\ell,k} \Fm_{\ell,k}^\herm \widetilde{\zv}^{(t)}_{\ell}.   \label{chest1}
% \end{align}
When $\Fm_{\ell,k}$ and $\Fm_{\ell,i}$ are nearly mutually orthogonal, i.e. $\Fm_{\ell,k}^\herm \Fm_{\ell,i} \approx \zerov$,
the subspace projection is able to significantly reduce the pilot contamination effect. Since the subspace estimation scheme in \cite{goettsch2022} achieves virtually the same performance as perfect channel subspace knowledge, we assume channel subspaces to be known in this paper.

\subsection{Uplink data rates with fronthaul quantization}
Let $s_k^{\rm ul}$ be the UL data symbol of UE $k$ (i.i.d. with mean zero and unit variance) at a given channel use of a generic RB. 
In this work, we consider ``smart RUs'' that operate according to O-RAN split option 7.2x in the upstream direction. 
More specifically, our chosen split option is most similar to 7.2x Cat-B ULPI-A, where RUs do channel estimation and equalization locally~\cite{ericsson2023ulpi}.
As in \cite{goettsch2022, li2023joint}, we consider ``local detection with cluster-level combining'', so that the equalized symbol (i.e., a local observation $r^{\rm ul}_{\ell,k}$ of $s^{\rm ul}_k$) at RU $\ell$ for each UE $k \in \Uc_\ell$ is sent to the cluster processor of $\Cc_k$ located at some DU over the fronthaul. The cluster-level symbol estimate is then computed at the DU.

In particular, the received UL signal at RU $\ell$ is given by
\begin{equation} 
	\yv_\ell^{\rm ul} = \sqrt{\SNR}  \sum_{i\in \Kc} \hv_{\ell,i} s_i^{\rm ul} + \zv_\ell^{\rm ul} ,  \label{ULchannel-RU}
\end{equation}
where $\zv_\ell^{\rm ul}$ is AWGN with components $\sim \Cc\Nc(0,1)$.
The local observation is obtained as a linear projection of $\yv_\ell^{\rm ul}$ onto a suitably defined receiver vector
$\vv_{\ell,k}$, such that 
%%% NOW INLINE %%%
% $r^{\rm ul}_{\ell,k} = \vv_{\ell,k}^\herm \yv_\ell^{\rm ul} .$
\begin{equation}
	r^{\rm ul}_{\ell,k} = \vv_{\ell,k}^\herm \yv_\ell^{\rm ul} . \label{local-receiver}
\end{equation} 
The receiver vectors $\{\vv_{\ell,k} : k \in \Uc_\ell\}$ are computed at RU $\ell$ using the local channel estimates
$\{\widehat{\hv}_{\ell,k} : k \in \Uc_\ell \}$. We consider linear MMSE receivers based on the partial local CSI at RU $\ell$, given by
\begin{equation} 
	\vv_{\ell,k} = \left ( \nu_\ell \Id + \SNR \sum_{i \in \Uc_{\ell}} \widehat{\hv}_{\ell,i} \widehat{\hv}_{\ell,i}^\herm \right )^{-1} \widehat{\hv}_{\ell,k} ,  \label{eq:lmmse}
\end{equation}
where $\nu_{\ell} \eqdef  1 + \SNR \sum_{i \neq \Uc_{\ell}}  \beta_{\ell,i}$ accounts for unknown interference and noise \cite{goettsch2022}.
The cluster processor for user $k$ collects the local observations from all RUs $\ell \in \Cc_k$ and forms
a cluster-level combined observation which is then passed to  the channel decoder for user $k$ as the soft-output of a virtual single user additive noise channel.

Since the fronthaul network has finite capacity links, we let each RU quantize its local observation $r^{\rm ul}_{\ell,k}$
with $B_{\ell,k}$ bits per sample before sending it to the cluster processor. We use the quantization scheme from \cite{li2023joint} based on rate-distortion theory, where $B_{\ell,k}$ depends on the signal strength at RU $\ell$ regarding UE $k$ defined as $\sigma_{\ell,k}^2 \defeq \EE[ |r_{\ell,k}^{\rm ul}|^2]$.
For a given desired distortion level $D$, each RU $\ell \in \Cc_k$ uses quantization rate
\begin{equation} 
	B_{\ell,k}  = \max \left \{ \log_2 \frac{\sigma_{\ell,k}^2}{D}, 0 \right \} \label{rateQ}
\end{equation}
to send the $r_{\ell,k}^{\rm ul}$ to the corresponding cluster processor.
Note that if $\sigma^2_{\ell,k} < D$ for some $(\ell,k) \in \Ec_{\rm ran}$, then $r_{\ell,k}^{\rm ul}$ is simply not sent from the RU to the cluster processor since the quantization rate is $B_{\ell,k} = 0$. This is equivalent to removing RU $\ell$ from cluster $\Cc_k$. 
The quantized local observation is given by \cite{li2023joint}:
\begin{equation} 
	\widehat{r}_{\ell,k}^{\rm ul} = \alpha_{\ell,k} r_{\ell,k}^{\rm ul}  + e_{\ell,k}, \label{bussgang_deco}
\end{equation}
where 
\begin{equation} 
	\alpha_{\ell,k} =  \frac{\sigma_{\ell,k}^2 - D}{\sigma_{\ell,k}^2}, \label{alpha_bussgang}
\end{equation}
and where $e_{\ell,k}$ is a zero-mean Gaussian random variable uncorrelated with $r_{\ell,k}^{\rm ul}$ and with variance 
\begin{equation}
	\widehat{\sigma}^2_{\ell,k} = (1 - D/\sigma_{\ell,k}^2) D.  \label{error_bussgang}
\end{equation} 
We let $\widehat{\rv}_k^{\rm ul} \in \CC^{|\Cc_k| \times 1}$ denote the vector of the quantized local observations $\{\widehat{r}^{\rm ul}_{\ell,k} : \ell \in \Cc_k\}$. The cluster-level combined observation is then given by
\begin{equation} 
	r_k^{\rm ul} = \wv_k^\herm \; \widehat{\rv}_k^{\rm ul}, \label{cluster-combining}
\end{equation}
where $\wv_k$ with elements $\left\{ w_{\ell,k} : \ell \in \Cc_k \right\}$ is the cluster-level combining vector that weighs the different local observations. It is computed to maximize the 
\textit{nominal} Signal to Interference plus Noise Ratio (SINR) of the channel with input $s_k^{\rm ul}$ and output
$r_k^{\rm ul}$, given the local knowledge of the cluster processor $\Cc_k$. Due to the limited local knowledge of cluster processor $\Cc_k$, we differentiate between the actual and nominal SINR. The nominal SINR is only used to compute $\wv_k$, while the actual SINR is used to compute the actual achievable physical layer user rate.
The maximization of the nominal SINR
with respect to the vector of combining coefficients $\wv_k$ is a classical generalized Rayleigh quotient maximization 
problem and is given in \cite{li2023joint}.

% We conclude this section by detailing the actual SINR and the resulting PHY per-user UL rate. 
% 	Writing explicitly the expression of $r^{\rm ul}_k$ from 
% 	(\ref{ULchannel-total}), (\ref{local-receiver}), (\ref{bussgang_deco}), and (\ref{cluster-combining}),
% 	we obtain~\footnote{Note that this expression is different from what obtained by the nominal model 
% 		(\ref{bussgang-vector-nominal}). However, while (\ref{bussgang-vector-nominal}) can be computed by the 
% 		$k$-th cluster processor,  (\ref{suca}) cannot, since it contains some unknown variables.} 
% 	\begin{eqnarray}
% 		r^{\rm ul}_k  & = & \sum_{\ell \in \Cc_k} w_{\ell,k}^* \widehat{r}^{\rm ul}_{\ell,k} \label{suca} \\
% 		& = & \sqrt{\SNR} \left ( \sum_{\ell \in \Cc_k} w_{\ell,k}^* \alpha_{\ell,k} \vv_{\ell,k}^\herm \hv_{\ell,k} \right ) s_k^{\rm ul} \label{useful-term} \\
% 		& & + \sqrt{\SNR} \sum_{i \neq k} \left ( \sum_{\ell \in \Cc_k} w^*_{\ell,k} \alpha_{\ell,k} \vv_{\ell,k}^\herm \hv_{\ell,i} \right ) s^{\rm ul}_i \label{multiuser-interference} \\
% 		& & +  \sum_{\ell \in \Cc_k} w^*_{\ell,k} \left ( \alpha_{\ell,k} \vv_{\ell,k}^\herm \zv_\ell^{\rm ul} + e_{\ell,k} \right ), 
% 		\label{channel-and-quantization-noise}
% 	\end{eqnarray}
% 	where (\ref{useful-term}) is the useful signal term, (\ref{multiuser-interference}) is the multiuser interference term, 
% 	and (\ref{channel-and-quantization-noise}) is the channel and quantization noise term.

The resulting {\em actual} SINR conditioned on the realization of all the channel vectors is given by \cite{li2023joint}
\begin{equation} 
    \SINR^{\rm ul}_k  = \frac{  \SNR \left |  \sum_{\ell \in \Cc_k} \widetilde{g}_{\ell, k, k} \right |^2 }
    {\sum_{\ell \in \Cc_k}  \widetilde{d}_{\ell,k}
    + \SNR \sum_{i \neq k} \left |  \sum_{\ell \in \Cc_k} \widetilde{g}_{\ell, k, i}  \right |^2 } ,  \nonumber
\end{equation}
where 
$\widetilde{g}_{\ell, k, i} \eqdef w^*_{\ell,k} \alpha_{\ell,k} \vv_{\ell,k}^\herm \hv_{\ell,i} $ and $\widetilde{d}_{\ell,k} \eqdef |w_{\ell,k}|^2 \left ( \alpha_{\ell,k}^2 \|\vv_{\ell,k}\|^2 + \widehat{\sigma}^2_{\ell,k} \right )$.
%
%
% \begin{align*}
% &\widetilde{g}_{\ell, k, i} \eqdef w^*_{\ell,k} \alpha_{\ell,k} \vv_{\ell,k}^\herm \hv_{\ell,i} \\
% \text{and \ } &\widetilde{d}_{\ell,k} \eqdef |w_{\ell,k}|^2 \left ( \alpha_{\ell,k}^2 \|\vv_{\ell,k}\|^2 + \widehat{\sigma}^2_{\ell,k} \right ).
% \end{align*}
As a performance measure of the physical layer,  we consider the UL {\em Optimistic Ergodic Rate} (OER) 
\cite{caire2018} given by 
\begin{eqnarray}
    R^{\rm ul}_k = \EE [ \log (1 + \SINR^{\rm ul}_k) ], \label{ergodic_rate_ul}
\end{eqnarray}
where the expectation is with respect to the small-scale fading, for given 
values of the LSFCs that depend on the placement of UEs and RUs, the pathloss function 
and cluster formation.

\subsection{Downlink data rates with fronthaul quantization} \label{DL_rates}
In the downlink (DL), the RUs operate according to split option 7.3 of 3GPP (e.g., see \cite{sup6620205g}), where the cluster processor sends the information bits to the RU, which carries out the modulation and precoding of the data symbols. Since each RU $\ell \in \Cc_k$ sends the same signal to UE $k$, the cluster processor sends the same information bits over the fronthaul to the RUs. The DL fronthaul traffic is thus of type multiple-multicast, requiring routers that enable multicast routing.
Since the information bits are sent, no quantization is necessary.\footnote{However, note that the choice of $D$ still affects the DL physical layer rates.
If $\sigma^2_{\ell,k} < D$ for some $(\ell,k) \in \Ec_{\rm ran}$, the corresponding RU is removed from $\Cc_k$.}

We use the approximate UL-DL reciprocity of \cite{goettsch2022, li2023joint} for the DL precoding vectors $\uv_{\ell,k}$. We let
$\uv_{\ell,k} \propto w^0_{\ell,k} \vv_{\ell,k}$ where $\vv_{\ell,k}$ are the local linear MMSE combiners for the UL defined in (\ref{eq:lmmse}) and $w^0_{\ell,k}$ are the cluster-level combining coefficients as in \eqref{cluster-combining} for the case of 
zero quantization distortion. 
% Then, the actual local precoding vectors $\uu_k$ are obtained by 
% stacking the blocks $w^0_{\ell,k} \vv_{\ell,k}$ for all $\ell \in \Cc_k$ and all-zero blocks for $\ell \neq \Cc_k$, and normalizing the 
% resulting vector of dimension $LM \times 1$ to have unit norm. 
The actual local precoding vectors $\uu_k$ are obtained by 
stacking the blocks $w^0_{\ell,k} \vv_{\ell,k}$ for all $\ell \in \Cc_k$ and all-zero blocks for $\ell \neq \Cc_k$, and normalizing the 
resulting vector of dimension $LM \times 1$ to have unit norm.
In this way, except for the common normalization factor, 
$\uv_{\ell,k}$ can be calculated from the local CSI at RU $\ell$, with sufficiently high resolution finite arithmetic. 

We let $\hh_k$ denote the $k$-th column of the overall channel matrix $\HH$ and $s^{\rm dl}_k$ the (coded) information symbol for UE $k$ (independent with mean zero and variance $q_k \geq 0$). 
The received signal sample at UE $k$ corresponding to one DL channel use is given by 
\begin{eqnarray} 
    y_k^{\rm dl} = \hh_k^\herm \uu_k s^{\rm dl}_k   + \sum_{j \neq k} \hh_k^\herm \uu_j s^{\rm dl}_j  + z^{\rm dl}_k , 
    \label{DLchannel}
\end{eqnarray}
%%% NOW INLINE %%%
% $
%     y_k^{\rm dl} = \hh_k^\herm \uu_k s^{\rm dl}_k   + \sum_{j \neq k} \hh_k^\herm \uu_j s^{\rm dl}_j  + z^{\rm dl}_k , 
%     \label{DLchannel}
% $
where, without loss of generality, we scale the received signal such that the 
noise is $z_k^{\rm dl} \sim \Cc\Nc(0, \SNR^{-1})$. 
The corresponding DL SINR is given by \cite{li2023joint}
\begin{eqnarray}
    \SINR^{\rm dl}_k & = & \frac{|\hh_k^\herm \uu_k|^2 q_k}{\SNR^{-1} + \sum_{j\neq k}   |\hh_k^\herm \uu_j|^2 q_j }. \label{DL-SINR} 
\end{eqnarray}
We choose uniform power allocation to all data streams, i.e., $q_k = 1$ for all $k \in \Kc$. 
The corresponding DL OER is given by 
\begin{eqnarray}
    R^{\rm dl}_k = \EE [ \log (1 + \SINR^{\rm dl}_k) ]. \label{ergodic_rate_dl}
\end{eqnarray}
Since the information bits are directly sent from the cluster processor located at some DU to the RUs, the fronthaul load corresponding to the DL is equal to the physical layer user rate, i.e., $R^{\rm dl}_k$.
%%%
%%%
%%%
%%%
%%%
\subsection{Remarks on the fronthaul load} 
The choice of the split options aims to reduce the fronthaul load by allocating many PHY functions to the RUs, thereby limiting the DU to cluster-level signal processing. We notice that the fronthaul load in this paper only considers quantities directly related to combining and detection in the UL and the information bits in the DL. The coefficients needed to compute the weights $\{ w_{\ell,k}\}$ and $\{ w_{\ell,k}^0 \}$ for UL combining and DL precoding, respectively, are not considered.
Note that these coefficients are constant during a coherence block of $T$ signal dimensions and that $T - \tau_p$ channel uses are dedicated to UL and DL data transmission. 
Assuming $\tau_p \approx 25$ and $T$ typically in the hundreds or thousands (see \cite{torres2021lower}), 
the fronthaul load related to the $T - \tau_p$ channel uses for data transmission is relatively large compared to the number of coefficients needed to compute the UL combining and DL precoding weights (see \cite{li2023joint}). We assume that these coefficients can be quantized very accurately without increasing significantly the fronthaul load. Therefore, we do not take into account the fronthaul traffic to exchange the coefficients for computing $\{ w_{\ell,k}\}$ and $\{ w_{\ell,k}^0 \}$ when defining the UL and DL fronthaul load per RB.

%%%%%%
%%%%%%
%%%%%%
%%%%%%
%%%%%%
%%%%%%
\section{Problem Statement}
In this section, we will describe how the user PHY data rates translate to the UL and DL fronthaul load of the whole system. 
The network operates in time division duplex mode, where ${ {\gamma_{\rm DL}}} \in (0,1)$ denotes the resource fraction allocated to the DL, and as a consequence $(1 - { {\gamma_{\rm DL}}})$ is allocated to the UL.
We use half-duplex fronthaul links as in \cite{li2023joint}, i.e., the fronthaul flow constraints must incorporate the fact that 
the data rate generated by the RUs in the UL is weighted by a factor $(1 - { {\gamma_{\rm DL}}})$ and the data rate
generated by the DUs in the DL is weighted by a factor ${ {\gamma_{\rm DL}}}$. 
The amount of UL fronthaul load transmitted from RU $\ell$ to router $q$, from router $q$ to router $q'$, and from 
router $q$ to DU $n$ regarding user $k$ is denoted 
as $x_{k}^{\text{ru}}(\ell, q)$, $x_{k}^{\text{fh}}(q,q')$, and $x_{k}^{\text{du}}(q,n)$, respectively. For the DL, we use a similar notation, where $y_{k}^{\text{ru}}(q, \ell)$, $y_{k}^{\text{fh}}(q,q')$ and $y_{k}^{\text{du}}(n, q)$ is the amount of DL fronthaul data with respect to user $k$ sent from router $q$ to RU $\ell$, from router $q$ to router $q'$, and from DU $n$ to router $q$, respectively. 
The load of any fronthaul link $(a,b) \in \Ec_{\rm front}$ is given by $\sum_{k} x_{k}(a,b)$, where 
$x_{k}(a,b)$ is one of the load variables previously introduced. 

\subsection{UL fronthaul load flow}
As defined in \eqref{rateQ}, RU $\ell \in \Cc_k$ generates a quantization rate of
$\fhratekUL$ bits per channel use that must be sent via the fronthaul to the DU $n$ that is the cluster processor of the user-centric cluster $\Cc_k$. 
Since the UL fronthaul contains only unicast flows, for each user $k$ every router $q$ must satisfy
\begin{align} \label{eq}
	\sum_{\ell}x_{k}^{\text{ru}}(\ell, q)+\sum_{q''}x_{k}^{\text{fh}}(q'',q)   =   \sum_{n} x_{k}^{\text{du}}(q,n)	+\sum_{q'}x_{k}^{\text{fh}}(q,q') .
\end{align}
Considering the TDD resource allocation fraction ${ {\gamma_{\rm DL}}}$, the UL fronthaul flow constraint for UE-RU pair $(\ell, k)$ is  
\begin{align}
	\sum_{q} x_{k}^{\text{ru}}(\ell, q) \geq a_{\ell, k} (1 - { {\gamma_{\rm DL}}}) \fhratekUL , \; \forall k,\ell , \label{eq_ul_k_to_q_sum}
\\
\text{and }	 x_{k}^{\text{ru}}(\ell, q) \leq a_{\ell, k} (1 - { {\gamma_{\rm DL}}}) \fhratekUL , \; \forall k,\ell,q,   \label{eq_ul_k_to_q}
\end{align}
where the UE-RU association binary variable $a_{\ell, k} = 1 $ if $(\ell, k) \in \Ec_{\rm ran}$, and $a_{\ell, k} = 0$ if $(\ell, k) \notin \Ec_{\rm ran}$.
% \begin{align}
% 	a_{\ell, k} = \begin{cases}
% 		1,& \text{if } (\ell, k) \in \Ec_{\rm ran}, \\
% 		0 ,& \text{if } (\ell, k) \notin \Ec_{\rm ran} . 
% 	\end{cases}   \nonumber
% \end{align}
Now, let $b_{k,n} \in \{0,1\}$ denote the cluster processor placement variable, defined by 
\begin{align}
	b_{k, n} = \begin{cases}
		1 ,& \text{if  $\Cc_k$ is hosted by DU $n$},  \\
		0 ,&  \text{otherwise}.  
	\end{cases}   \nonumber
\end{align}
We have the constraints that each $\Cc_k$ must be hosted by exactly one DU and a computation capacity constraint per DU, i.e., 
\begin{gather}
	\sum_{n=1}^{N} b_{k,n}  =  1 , \; \forall k,  \label{sumb} \\
	\sum_{k = 1}^K b_{k,n} \leq Z_n,  \; \forall n ,
\end{gather}
where $Z_n$ is a computation limit for the number of cluster processors at any DU $n$. Given the cluster processor placement, we note that the received flow relative to 
user $k$ to DU $n$ hosting $\Cc_k$ must be not smaller than the sum of source rates 
$(1 - { {\gamma_{\rm DL}}}) \fhratekUL$ over all RUs $\ell \in \Cc_k$, i.e., 
\begin{align}\label{d3}
	\sum_{q} x_{k}^{\text{du}}(q,n)\geq b_{k,n} (1 - { {\gamma_{\rm DL}}}) \sum_{\ell \in \Cc_k} \fhratekUL , \; \forall k, n .
\end{align}
We also introduce the 
individual load variables upper bounds
\begin{align}\label{d4}
	x_{k}^{\text{du}}(q,n)\leq b_{k,n} (1 - { {\gamma_{\rm DL}}}) \sum_{\ell \in \Cc_k}  \fhratekUL , \; \forall k, q, n .
\end{align}
In particular, this means that if $b_{k,n} = 0$, the rate
relative to user $k$ from any connected router $q$ to $n$ will be zero. 

\subsection{DL fronthaul load flow}

As explained in Section \ref{DL_rates}, the DL fronthaul traffic is of type multiple multicast and the number of information bits per channel use necessary 
to encode the DL signal for user $k$ at each RU $\ell \in \Cc_k$ is equal to the DL PHY rate $\fhratekDL$.
Since the DL fronthaul traffic is of multicast type, the flow conservation at the routers no longer applies (e.g., 
intermediate nodes may duplicate some input to multiple output links). 
Considering the data of user $k$, the output data of a router $q$ to any RU $\ell$ or router $q''$ must be less than or equal to the sum input data from the cluster processor and other routers, i.e.,
\begin{align}\label{dl1}
	\sum_{n} y_{k}^{\text{du}}(n,q)+\sum_{q'}y_{k}^{\text{fh}}(q',q)\geq y_{k}^{\text{ru}}(q,\ell), \; \forall k, q, \ell, 
\end{align}
and 
\begin{align}\label{dl2}
	\sum_{n} y_{k}^{\text{du}}(n,q)+\sum_{q'}y_{k}^{\text{fh}}(q',q)\geq y_{k}^{\text{fh}}(q,q''), \; \forall k, q, q'' .
\end{align}
In the DL, the DU hosting $\Cc_k$ must transmit at least $\fhratekDL$ bits per channel use to the connected routers, i.e., 
\begin{align}\label{dl3}
	\sum_{q} y^{\text{du}}_{k}(n,q) \geq b_{k,n} { {\gamma_{\rm DL}}} \fhratekDL, \; \forall k,n.
\end{align}
On each individual link to a router $q$, the DU needs to transmit at most $\fhratekDL$ bits and only the DU hosting cluster $\Cc_k$  transmits fronthaul data for user $k$. These two constraints are summarized as 
\begin{align}\label{dl4}
	y^{\text{du}}_{k}(n,q) \leq b_{k,n} { {\gamma_{\rm DL}}} \fhratekDL, \; \forall k,n,q.
\end{align}
The constraint that guarantees that each RU $\ell \in \Cc_k$ receives $\fhratekDL$ fronthaul bits for user $k$ is formulated as
\begin{align}\label{dl5}
	\sum_{q} y_{k}^{\text{ru}}(q, \ell) \geq a_{k,\ell} { {\gamma_{\rm DL}}} \fhratekDL, \; \forall k,\ell.
\end{align}

%%%%%%%%%%%%%%%%%%%%%%%%%%%%%%%%%
\subsection{Fronthaul optimization problem}   \label{jointopt}
Let $\Cc = \left\{ C_L, C_Q, C_D \right\}$ the
maximum link loads for RU-router, router-router, and router-DU links (with corresponding weights $\eta_{L/Q/D}$).
Further, $\Bc$, $\Xc$ and $\Yc$ denote the ensemble of all $\{b_{k,n} : \forall k, n\}$, UL load and DL load variables, respectively. 
Then, we can use the mixed-integer linear program (MILP) of \cite{li2023joint} for joint optimization:
% Then, 
% the joint optimization problem is formulated as a  mixed-integer linear program (MILP) by \cite{li2023joint}
\begin{subequations}\label{opt_problem_hd}
    \begin{eqnarray}
        &\min\limits_{\Bc, \Cc, \Xc, \Yc} & \eta_L C_L + \eta_Q C_Q + \eta_D C_D \hspace{3.5cm} \\
        &\text{s. t.}
        & \sum_{k}  \left( x^{\text{ru}}_{k}(\ell, q) + y_{k}^{\text{ru}}(q,\ell)  \right)		\leq C_L, \; \forall \ell, q, \label{HD1}\\
        &&\sum_{k}  \left( x_{k}^{\text{fh}}(q,q') + y_{k}^{\text{fh}}(q,q')  \right)			\leq C_Q, \; \forall q, q',  \\
        &&\sum_{k}  \left( x_{k}^{\text{du}}(q,n) + y_{k}^{\text{du}}(n,q)  \right)		\leq C_D, \; \forall q, n, \label{HD2}\\	
        &&  \eqref{eq}-\eqref{dl5},
    \end{eqnarray}
\end{subequations}
where constraints \eqref{HD1}-\eqref{HD2} ensure that each link load is less than the link capacity. Note that MILPs can be solved by existing highly efficient optimization tools \cite{elie2019}. To implement \eqref{opt_problem_hd}, we utilize the MATLAB function intlinprog.

\section{Numerical Results}
We start this section with an overview of the considered network and parameters, before showing the impact of different $K$ and $D$ on the PHY SE and fronthaul load. 
% The results are briefly discussed in terms of practical network deployments and implications for user scheduling.
Our cell-free massive MIMO network has an area of $200 \times 200 \text{ m}^2$ with a torus topology to avoid boundary effects. The LSFCs are given by the 3GPP urban microcell pathloss model from \cite[Table 7.4.2-1]{3gpp38901}. The spatial correlation between antennas follows the one-ring scattering model
with angular support 
$\Theta_{\ell,k} = [\theta_{\ell,k} - \Delta/2, \theta_{\ell,k} + \Delta/2]$, i.e., it is centered at angle $\theta_{\ell,k}$ of the LOS
between RU $\ell$ and UE $k$ with angular spread $\Delta$. 
Then, the channel between UE $k$ and RU $\ell$ is given by
%%% NOW INLINE %%%
% $\hv_{\ell,k} = \sqrt{\frac{\beta_{\ell,k} M}{|\Sc_{\ell,k}|}}  \Fm_{\ell,k} \nuv_{\ell, k}$,
\begin{align*} 
\hv_{\ell,k} = \sqrt{\frac{\beta_{\ell,k} M}{|\Sc_{\ell,k}|}}  \Fm_{\ell,k} \nuv_{\ell, k},  %\label{channel_model}
\end{align*}
where the index set  $\Sc_{\ell,k} \subseteq \{0,\ldots, M-1\}$ includes all integers $m$ such that $2\pi m/M \in \Theta_{\ell,k}$
(where angles are taken modulo $2\pi$), 
$\Fm_{\ell,k}$ is the submatrix extracted from the $M \times M$ unitary DFT matrix $\Fm$ by taking the columns indexed by $\Sc_{\ell,k}$,  
and  $\nuv_{\ell,k} \in \CC^{|\Sc_{\ell,k}| \times 1}$ has i.i.d. components 
$\sim \Cc\Nc(0,1)$. Hence,  $\hv_{\ell,k}$ is a Gaussian zero-mean random vector confined in the subspace spanned by the columns 
of $\Fm_{\ell,k}$.  

\subsection{Simulation setup}
We consider a system with $L=20$ RUs, each equipped with $M = 10$ antennas and placed on a rectangular $5 \times 4$ grid. 
We set $\Delta = \pi/8$, $\taudmrs = 25$, the maximum cluster size $|\Cc_k| \leq \Cc_{\rm max} = 7$, 
and the SNR threshold $\eta = 1$ in (\ref{eq:snr_threshold}).
We define the expected pathloss $\bar{\beta}$ at distance $2.5 d_L$ between UE and RU, where $d_L = \sqrt{\frac{A}{\pi L}}$ is the radius of a disk of area equal to $A/L$.
The UL energy per symbol of each UE is $\bar{\beta} M \SNR = 1$ (i.e., 0 dB) and thus depends on the RU density and the number of RU antennas. This choice leads to a certain level of coverage overlap among RUs, such that each UE is likely to associate with several RUs.
The complex baseband noise power 
density is $N_0 = -174 \text{ dBm/Hz}$.
The RUs are partially connected to $Q=5$ routers, which in turn are partially connected to $N=4$ DUs.\footnote{We use the same fronthaul links as \cite{li2023joint} and refer the reader to \cite{li2023joint} for a description and illustration of what links exist between RUs, routers and DUs.} For each of the DUs, we impose $Z_n = K/2$.
 The optimization weights $\eta_L$, $\eta_Q$, $\eta_D$ are set to $1$.  

Each coherence block contains
$T=200$ signal dimensions, of which $\tau_p = 20$ are used for UL pilots and ${\gamma_{\rm DL}} = 0.8$.
We compute the expectation in \eqref{ergodic_rate_ul} and \eqref{ergodic_rate_dl} over $100$ channel realizations and
define the total UL/DL SE, i.e., the UL/DL PHY SE in bits per channel use (or bit/s/Hz) of all users, as ${\rm SE}^{\rm ul} = \left(1 - { {\gamma_{\rm DL}}} \right) \left( 1- \frac{\taudmrs}{T} \right) \sum_{k \in \Kc}  R^{\rm ul}_k$ and ${\rm SE}^{\rm dl} = { {\gamma_{\rm DL}}}  \left( 1- \frac{\taudmrs}{T} \right) \sum_{k \in \Kc}   R^{\rm dl}_k \; \text{bit/s/Hz}$.
% \begin{align}
% 	{\rm SE}^{\rm ul} &= \left(1 - { {\gamma_{\rm DL}}} \right) \left( 1- \frac{\taudmrs}{T} \right) \sum_{k \in \Kc}  R^{\rm ul}_k \; \; \; \text{bit/s/Hz} ,  \label{eq:sum_phy_rate}
% 	\\
% 	{\rm SE}^{\rm dl} &= { {\gamma_{\rm DL}}}  \left( 1- \frac{\taudmrs}{T} \right) \sum_{k \in \Kc}   R^{\rm dl}_k \; \; \; \text{bit/s/Hz} .\label{eq:sum_phy_rate-dl}
% \end{align}
% where $R^{\rm ul}_k$ and $R^{\rm dl}_k$ are given by \eqref{ergodic_rate_ul} and  \eqref{ergodic_rate_dl}, respectively. 
% The expectation in \eqref{ergodic_rate_ul} and \eqref{ergodic_rate_dl} is computed over $100$ channel realizations. 
The total sum PHY SE is given by ${\rm SE}_{\rm tot} = {\rm SE}^{\rm ul} + {\rm SE}^{\rm dl}$.

\subsection{Evaluation of the user load and quantization level}
We evaluate the impact of the user load $K$ and distortion level $D$ on the fronthaul load and PHY SE. 
We let $K = [75, 200]$, which ranges from a lightly loaded to an overloaded system compared to $LM$ (see \cite{gottsch2023fairness}). Recall that a UE-RU association is removed if $\sigma^2_{\ell,k} < D$. Therefore, the smallest distortion level $D$ is chosen such that $D/\sigma^2_{\rm min} < 1$, where $\sigma^2_{\rm min} = \min_{(\ell,k) \in \Ec_{\rm ran}}\sigma^2_{\ell,k}$.
% $\forall (\ell, k) \in \Ec_{\rm ran}$ for all considered $K$.
%%%
%%%
%%%
% Note that for larger $K$ the multi-user interference increases, so the values of $\{\sigma^2_{\ell,k}  : (\ell, k) \}$ become smaller.
% \footnote{Since for larger $K$ the multi-user interference increases, the values of $\{\sigma^2_{\ell,k}  : (\ell, k) \}$ become smaller as $K$ increases.}
%%%
%%%
%%%
As $D$ is increased, $B_{\ell,k}$ in \eqref{rateQ} decreases and more UE-RU associations are removed. 

\begin{figure}[t!]
	\centering
	\begin{subfigure}{0.49\linewidth}
		\includegraphics[width=\linewidth]{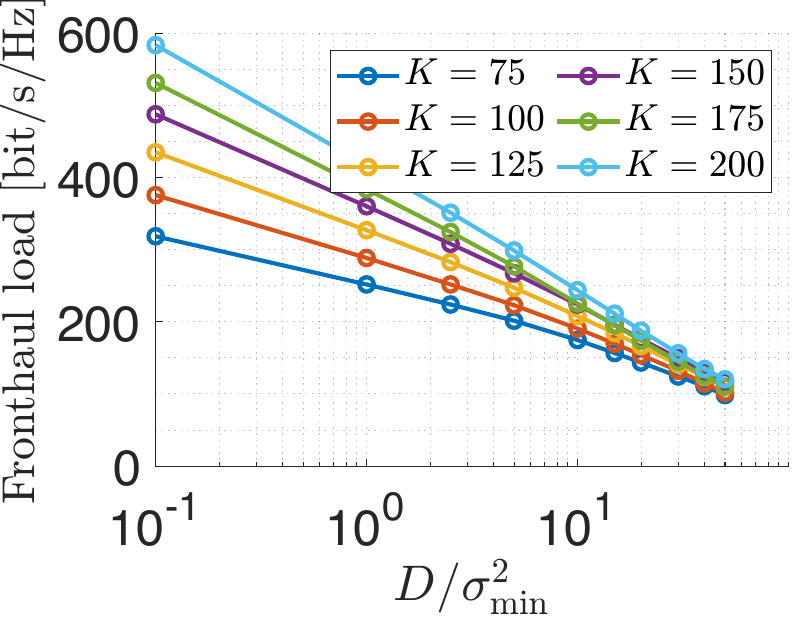}
	\end{subfigure}
	\begin{subfigure}{0.49\linewidth}
		\includegraphics[width=\linewidth]{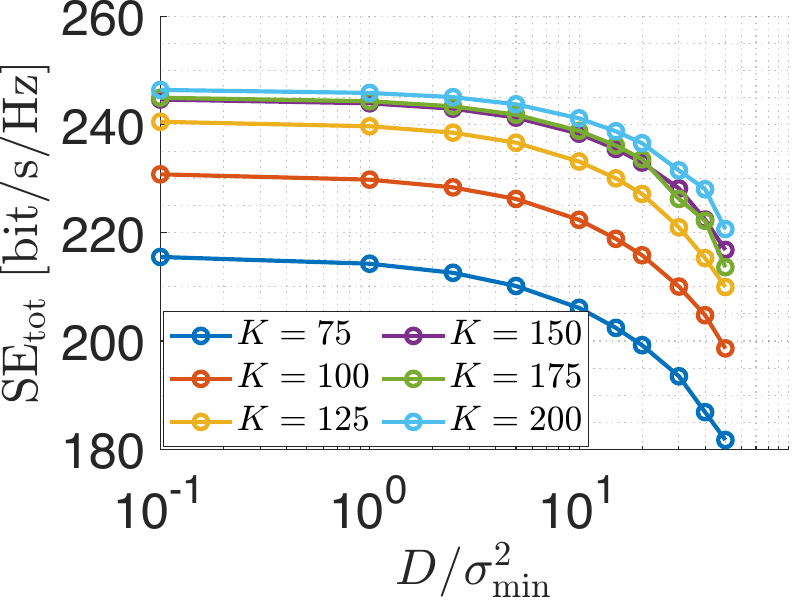}
	\end{subfigure}
    \vspace{-15pt}
	\caption{Sum UL/DL fronthaul load and SE vs. $D$. }%{ [UL/DL NOW IN SEPARATE PLOTS]}}
	\label{UL_DL_imb_PHY_vs_FH}
    \vspace{-10pt}
\end{figure}

\begin{figure}[t!]
	\centering
	\begin{subfigure}{0.49\linewidth}
		\includegraphics[width=\linewidth]{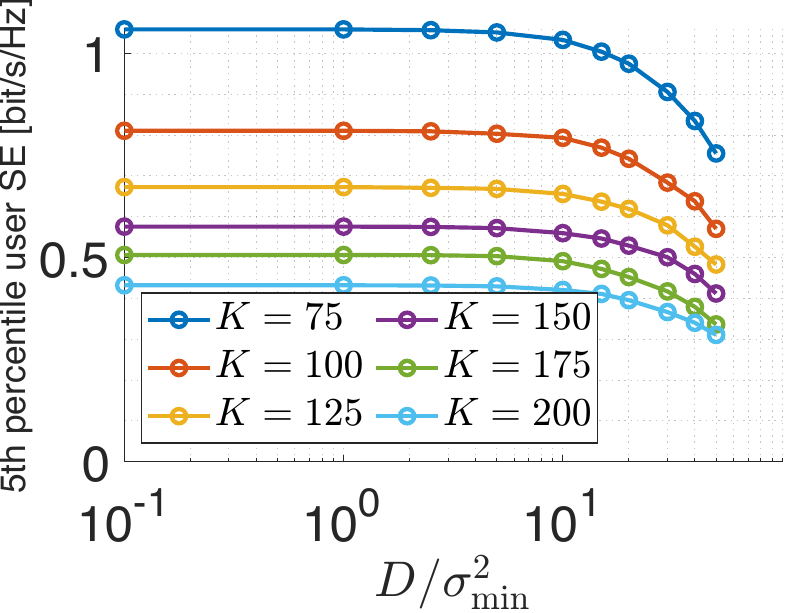}
	\end{subfigure}
	\begin{subfigure}{0.49\linewidth}
		\includegraphics[width=\linewidth]{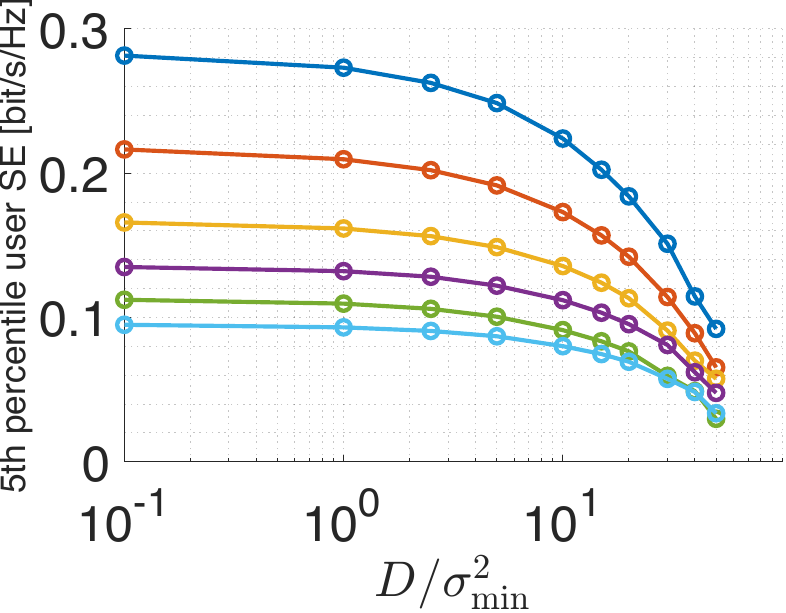}
	\end{subfigure}
    \vspace{-15pt}
	\caption{5th percentile user SE in the DL (left) and UL (right, same legend as for the UL applies) vs. $D$.}%{ [UL/DL NOW IN SEPARATE PLOTS]}}
	\label{UL_DL_pctile_phy}
    \vspace{-15pt}
\end{figure}

As Fig. \ref{UL_DL_imb_PHY_vs_FH} shows, each increase of $D$ leads to a significant reduction of the fronthaul load, i.e., $\eta_L C_L + \eta_Q C_Q + \eta_D C_D$, while the sum PHY SE only decreases for large $D$. 
We note that as $D$ is increased, the fronthaul load becomes smaller for large $K$ than for small $K$ in some cases. 
This is explained by smaller $\{\sigma^2_{\ell,k}  : (\ell, k) \}$ for large $K$ due to multi-user interference. Increasing $D$ leads to more UE-RU associations being removed and reduced quantization rates $B_{\ell,k}$ compared to a low user load.
% This occurs because $\{\sigma^2_{\ell,k}  : (\ell, k) \}$ are smaller for large $K$ due to multi-user interference. With a fixed $D$ and large $K$, this leads to more UE-RU associations being removed and smaller quantization rates $B_{\ell,k}$ compared to low user load.
When $D$ is increased in the high distortion regime, the sum PHY SE first decreases slightly and then, as more UE-RU associations are removed, significantly. Fig. \ref{UL_DL_pctile_phy} confirms and explains this behavior. The degradation of the 5th percentile DL user SE ${\gamma_{\rm DL}} \left( 1- \frac{\taudmrs}{T} \right) R^{\rm dl}_k$ is observed at $D/\sigma^2_{\rm min} \geq 10$. No quantization is needed in the DL, so this effect is caused solely by removing UE-RU associations. The degradation of $(1 -{\gamma_{\rm DL}}) \left( 1- \frac{\taudmrs}{T} \right) R^{\rm ul}_k$ in the UL occurs for smaller $D$ due to fronthaul quantization distortion. Since a larger fraction of resources is used in the DL, the impact of $D$ on ${\rm SE}_{\rm tot}$ in Fig. \ref{UL_DL_imb_PHY_vs_FH} becomes more significant for $D/\sigma^2_{\rm min} \geq 10$. 
We further notice that the average cluster size decreases from nearly $\Cc_{\rm max} = 7$ RUs with $D/\sigma^2_{\rm min} = 1$ to $\approx 6$ and $\approx 5$ RUs at $D/\sigma^2_{\rm min} = 10$ and $D/\sigma^2_{\rm min} = 20$, respectively. 
% At $D/\sigma^2_{\rm min} = 20$ and $D/\sigma^2_{\rm min} = 50$, where the SE decreases substantially, the average cluster size is $\approx 5$ and $\approx 4$ RUs, respectively. 
As another result from Fig. \ref{UL_DL_imb_PHY_vs_FH} we observe that the sum SE grows with larger $K$, but only in the range $K = [75, 150]$. For $K>150$, multi-user interference becomes more severe and 
% Serving $K=175$ UEs does not lead to a further significant gain, and $K=200$ even leads to a loss in sum SE. 
%Therefore, 
%from a PHY performance perspective, 
it is recommended not to serve much more than $150$ UEs.  

\subsection{Concluding remarks and future work}
A wide range of user loads can be supported by the same cell-free massive MIMO RAN and fronthaul network if the distortion level $D$ is optimized. 
% The quantization of local observations $\{ r^{\rm ul}_{\ell,k} \}$ at the RUs with \eqref{rateQ} is a principled ($B_{\ell.k}$ is directly related to the received signal strength $\sigma_{\ell,k}^2$) and practically applicable scheme.
For example, in this setup, $D/\sigma^2_{\rm min} \approx 5$ for $K = \{75, 100\}$ and $D/\sigma^2_{\rm min} \approx 10$ for $K = \{125, 150\}$ are good choices. Then, the fronthaul load for the recommended user load $K = [75, 150]$ is between $200$ and $225$ bit/s/Hz (i.e., in a relatively small margin) and the PHY SE is degraded only very slightly compared to smaller $D$. 
This is also an interesting result for scheduling problems. If the total number of users $\Ktot$ in the network area is varying, the scheduler can flexibly react to different user loads and demands, and change the number $K$ of simultaneously active users without overloading the fronthaul network.
We conclude that dense cell-free massive MIMO deployments are a very promising approach to achieve resilient 6G (radio access and fronthaul) networks with regard to varying user densities.

Future work includes investigating the impact of the distortion level for different fronthaul network topologies and fronthaul load optimization algorithms, and to study scheduling with flexible user loads under fronthaul constraints. 

\section{Acknowledgments}
The authors acknowledge the financial support by the Federal Ministry of Research, Technology and Space of Germany in the programme of “Souverän. Digital. Vernetzt.” Joint project 6G-RIC, project identification number 16KISK030.
%F. G\"ottsch, L. Miretti, G. Caire and S. Sta\'nczak acknowledge the financial support by the Federal Ministry of Education and Research of Germany in the programme of “Souverän. Digital. Vernetzt.” Joint project 6G-RIC, project identification number: 16KISK020K (L. Miretti and S. Sta\'nczak) and 16KISK030 (all mentioned authors).

\bibliographystyle{IEEEtran}
\bibliography{fronthaul_references}

\end{document}